\documentclass[12pt]{article}
\usepackage{amssymb}

\usepackage{graphicx}

\ifx\pdfoutput\undefined
    \relax
\else
    \usepackage{epstopdf}
\fi

\makeatletter
\def\numberbysection{\@addtoreset{equation}{section}
 	\def\theequation{\thesection.\arabic{equation}}}
\makeatother

\numberbysection

\newcommand{\be}{\begin{eqnarray}}
\newcommand{\ee}{\end{eqnarray}}
\newcommand{\non}{\nonumber}
\newcommand{\tr}{\mathop{\rm tr}\nolimits}
\newcommand{\id}{\mathbb{I}}
\newcommand{\ch}{\mathop{\rm cosh}\nolimits}
\newcommand{\sh}{\mathop{\rm sinh}\nolimits}
\newcommand{\tnh}{\mathop{\rm tanh}\nolimits}
\newcommand{\cth}{\mathop{\rm coth}\nolimits}
\newcommand{\csch}{\mathop{\rm cosech}\nolimits}
\newcommand{\sech}{\mathop{\rm sech}\nolimits}
\newcommand{\h}{\ensuremath{\mathsf{h}}}
\newcommand{\sgn}{\mathop{\rm sgn}\nolimits}

\begin{document}

\begin{titlepage}
\strut\hfill UMTG--249
\vspace{.5in}
\begin{center}

\LARGE Boundary energy of the open XXZ chain\\
\LARGE from new exact solutions\\[1.0in]
\large Rajan Murgan, Rafael I. Nepomechie and Chi Shi\\[0.8in]
\large Physics Department, P.O. Box 248046, University of Miami\\[0.2in]  
\large Coral Gables, FL 33124 USA\\

\end{center}

\vspace{.5in}

\begin{abstract}
Bethe Ansatz solutions of the open spin-${1\over 2}$ integrable XXZ
quantum spin chain at roots of unity with nondiagonal boundary terms
containing two free boundary parameters have recently been proposed.
We use these solutions to compute the boundary energy (surface energy)
in the thermodynamic limit.
\end{abstract}
\end{titlepage}

\setcounter{footnote}{0}

\noindent
{\em \large In memory of Daniel Arnaudon.} 

\section{Introduction}\label{sec:intro}

While the solution of the open spin-${1\over 2}$ XXZ quantum spin
chain with {\em diagonal} boundary terms has long been known \cite{Ga,
ABBBQ, Sk}, the solution of the general integrable case, with
the Hamiltonian \cite{dVGR,GZ}
\be
{\mathcal H} &=& {\mathcal H}_{0}
+ {1\over 2}\sh \eta \Big[ 
\cth \alpha_{-} \tnh \beta_{-}\sigma_{1}^{z}
+ \csch \alpha_{-} \sech \beta_{-}\big( 
\ch \theta_{-}\sigma_{1}^{x} 
+ i\sh \theta_{-}\sigma_{1}^{y} \big) \non \\
& & \quad -\cth \alpha_{+} \tnh \beta_{+} \sigma_{N}^{z}
+ \csch \alpha_{+} \sech \beta_{+}\big( 
\ch \theta_{+}\sigma_{N}^{x}
+ i\sh \theta_{+}\sigma_{N}^{y} \big)
\Big]  \,, \label{Hamiltonian} 
\ee
where
\be 
{\mathcal H}_{0} = {1\over 2}\sum_{n=1}^{N-1}\left( 
\sigma_{n}^{x}\sigma_{n+1}^{x}+\sigma_{n}^{y}\sigma_{n+1}^{y}
+\ch \eta\ \sigma_{n}^{z}\sigma_{n+1}^{z}\right) \,, 
\ee
(which contains also {\em nondiagonal} boundary terms) has remained
elusive.  Here $\sigma^{x} \,, \sigma^{y} \,, \sigma^{z}$ are the
standard Pauli matrices, $\eta$ is the bulk anisotropy parameter,
$\alpha_{\pm} \,, \beta_{\pm} \,, \theta_{\pm}$ are arbitrary boundary
parameters, and $N$ is the number of spins.

Progress has recently been made on this problem.  Indeed, a Bethe
Ansatz solution is now known \cite{CLSW, Ne, NR, YNZ} if the boundary
parameters obey the constraints
\be
\alpha_- + \epsilon_1 \beta_- + \epsilon_2 \alpha_+ + 
\epsilon_3 \beta_+ = \epsilon_0 (\theta_- -\theta_+) 
+\eta k + \frac{1-\epsilon_2}{2}i\pi \quad {\rm mod}\, (2i\pi) \,,
\quad \epsilon_1 \epsilon_2 \epsilon_3=+1 \,, 
\label{constraint}
\ee
where $\epsilon_i=\pm 1$, and $k$ is an integer such
that $|k|\leq N-1$ and $N-1+k$ is even. Finite-size effects for this
model and for the boundary sine-Gordon model \cite{GZ} have been 
computed on the basis of this solution \cite{AN, ABNPT}. (Related
results have been obtained by different methods in \cite{TBA}.)
Many interesting further applications and generalizations of this solution have
also been found (see, e.g., \cite{generalizations}).

Additional Bethe Ansatz solutions with up to two free boundary
parameters have been proposed in \cite{MN1, MN2}.  Completeness of
these new solutions is straightforward, in contrast to the case
(\ref{constraint}) \cite{NR}.  A noteworthy feature of the
solution \cite{MN2} is the appearance of a {\em generalized} $T-Q$ relation
of the form
\be
t(u)\ Q_{1}(u) &=& Q_{2}(u') + Q_{2}(u'') \,, \non \\
t(u)\ Q_{2}(u) &=& Q_{1}(u') + Q_{1}(u'') \,,
\label{generalizedTQ}
\ee
involving two $Q$-operators, instead of the usual one \cite{Ba}.
However, unlike the case (\ref{constraint}),
these new solutions hold only at roots of unity, i.e., for
bulk anisotropy values
\be
\eta = {i \pi\over p+1} \,,
\label{eta}
\ee 
where $p$ is a positive integer.

The aim of this paper is to use the new solutions \cite{MN1, MN2} to
investigate the ground state in the thermodynamic ($N \rightarrow
\infty$) limit.  For definiteness, we focus on two particular cases:
\be
\mbox{Case I: The bulk anisotropy parameter has values (\ref{eta}) 
with $p$ {\it even};} \non \\
\mbox{the boundary parameters $\beta_{\pm}$ are
arbitrary, and $\alpha_{\pm}=\eta$, $\theta_{\pm} = 0$ \cite{MN1}}
\label{caseI}
\ee 
\be
\mbox{Case II: The bulk anisotropy parameter has values (\ref{eta}) 
with $p$ {\it odd};} \non \\
\mbox{the boundary parameters $\alpha_{\pm}$ are
arbitrary, and $\beta_{\pm}=\theta_{\pm} = 0$ \cite{MN2}}
\label{caseII}
\ee 

\noindent 
We also henceforth restrict to even values of $N$. 
For each of these cases, we determine the density of Bethe roots
describing the ground state in the thermodynamic limit, for
suitable values of the boundary parameters; 
and we compute the corresponding boundary (surface) energies.
\footnote{For the case of diagonal boundary terms, the boundary energy
was first computed numerically in \cite{ABBBQ}, and then analytically in
\cite{HQB}.}
We find that the results coincide with the boundary energy computed in
\cite{AN} for the case (\ref{constraint}), namely, \footnote{Here we
correct the misprint in Eq.  (2.29) of \cite{AN}, as already noted in
\cite{ABNPT}.}
\be
E_{boundary}^{\pm} &=& - {\sin \mu\over 2\mu} 
\int_{-\infty}^{\infty} d\omega\ 
{1\over 2\cosh (\omega/ 2)}
\Big\{ 
{\sinh((\nu-2)\omega/4) \over 2\sinh(\nu \omega/4)} 
-{1\over 2} \non \\
&+& \sgn(2a_{\pm}-1){\sinh((\nu-|2a_{\pm}-1|)\omega/2) \over 
\sinh(\nu \omega/2)} + 
{\sinh(\omega/2) \cos (b_{\pm}\omega) \over 
\sinh(\nu \omega/2)} \Big\} \non \\
&+& {1\over 2} \sin \mu \cot (\mu a_{\pm}) 
 -{1\over 4}\cos \mu \,,
\label{ANboundenergy}
\ee
where 
\be
\mu = -i \eta =  {\pi\over p+1} \,, \quad 
\nu = {\pi\over \mu} = p+1 \,, \quad
\alpha_{\pm} = i \mu a_{\pm} \,, \quad 
\beta_{\pm} = \mu b_{\pm} \,,
\label{notation}
\ee 
and $\sgn(n) = {n\over |n|}$ for $n \ne 0$. 

The outline of this paper is as follows.  In Section
\ref{sec:transfer} we briefly review the transfer matrix and its
relation to the Hamiltonian (\ref{Hamiltonian}).  In Sections
\ref{sec:caseI} and \ref{sec:caseII} we treat Cases I (\ref{caseI})
and II (\ref{caseII}), respectively.  We conclude in Section
\ref{sec:discuss} with a brief discussion of our results.

\section{Transfer matrix}\label{sec:transfer}

The transfer matrix $t(u)$ of the model is given by \cite{Sk}
\be
t(u) = \tr_{0} K^{+}_{0}(u)\  
T_{0}(u)\  K^{-}_{0}(u)\ \hat T_{0}(u)\,,
\label{transfer}
\ee
where $T_{0}(u)$ and $\hat T_{0}(u)$ are the monodromy matrices 
\be
T_{0}(u) = R_{0N}(u) \cdots  R_{01}(u) \,,  \qquad 
\hat T_{0}(u) = R_{01}(u) \cdots  R_{0N}(u) \,,
\label{monodromy}
\ee
and $\tr_{0}$ denotes trace over the ``auxiliary space'' 0.
The $R$ matrix is given by
\be
R(u) = \left( \begin{array}{cccc}
	\sinh  (u + \eta) &0            &0           &0            \\
	0                 &\sinh  u     &\sinh \eta  &0            \\
	0                 &\sinh \eta   &\sinh  u    &0            \\
	0                 &0            &0           &\sinh  (u + \eta)
\end{array} \right) \,,
\label{bulkRmatrix}
\ee 
where $\eta$ is the bulk anisotropy parameter; and $K^{\mp}(u)$ are
$2 \times 2$ matrices whose components
are given by \cite{dVGR, GZ}
\be
K_{11}^{-}(u) &=& 2 \left( \sinh \alpha_{-} \cosh \beta_{-} \cosh u +
\cosh \alpha_{-} \sinh \beta_{-} \sinh u \right) \non \\
K_{22}^{-}(u) &=& 2 \left( \sinh \alpha_{-} \cosh \beta_{-} \cosh u -
\cosh \alpha_{-} \sinh \beta_{-} \sinh u \right) \non \\
K_{12}^{-}(u) &=& e^{\theta_{-}} \sinh  2u \,, \qquad 
K_{21}^{-}(u) = e^{-\theta_{-}} \sinh  2u \,,
\label{Kminuscomponents}
\ee
and
\be
K^{+}(u) = \left. K^{-}(-u-\eta)\right\vert_{(\alpha_-,\beta_-,\theta_-)\rightarrow
(-\alpha_+,-\beta_+,\theta_+)} \,,
\ee 
where $\alpha_{\mp} \,, \beta_{\mp} \,, \theta_{\mp}$ are the boundary
parameters.

For $u=0$, the transfer matrix is given by
\be
t(0) = c_{0} \id  \,, \qquad
c_{0} =  -8 \sh^{2N}\eta \ch \eta \sh \alpha_{-} \sh \alpha_{+} 
\ch \beta_{-} \ch \beta_{+} \,.
\label{initial}
\ee
For $\eta \ne i\pi/2$, the Hamiltonian (\ref{Hamiltonian}) is related 
to the first derivative of the transfer matrix at $u=0$,
\be
{\mathcal H} = c_{1} t'(0) + c_{2} \id \,,
\label{Htrelation}
\ee
where
\be
c_{1} &=& -\left( 16 \sinh^{2N-1} \eta \cosh \eta
\sinh \alpha_{-} \sinh \alpha_{+}
\cosh \beta_{-} \cosh \beta_{+} \right)^{-1} \,, \non \\
c_{2} &=& - {\sinh^{2}\eta  + N \cosh^{2}\eta\over 2 \cosh \eta} 
\,.
\ee 
For the special case $\eta = i\pi/2$ (i.e., $p=1$),
\be
t'(0) = d_{0} \id \,, \qquad 
d_{0} = (-1)^{N+1} 8 i \sh \alpha_{-} \sh \alpha_{+} 
\ch \beta_{-} \ch \beta_{+} \,, 
\label{initialp1}
\ee
and the Hamiltonian (\ref{Hamiltonian}) is related 
to the second derivative of the transfer matrix at $u=0$ \cite{Ne01},
\be
{\mathcal H} = d_{1} t''(0) \,, \qquad
d_{1} = (-1)^{N+1} \left( 32 \sh \alpha_{-} \sh \alpha_{+} 
\ch \beta_{-} \ch \beta_{+} \right)^{-1} \,.
\label{Htrelationp1}
\ee

\section{Case I: $p$ even}\label{sec:caseI}

For Case I (\ref{caseI}), the Hamiltonian (\ref{Hamiltonian}) becomes
\be
{\mathcal H} = {\mathcal H}_{0}
+ {1\over 2} \left( 
\ch \eta \tnh \beta_{-}\sigma_{1}^{z}
+ \sech \beta_{-} \sigma_{1}^{x} 
-\ch \eta \tnh \beta_{+} \sigma_{N}^{z}
+ \sech \beta_{+} \sigma_{N}^{x} \right)  \,, 
\label{HamiltonianI} 
\ee
which is Hermitian for $\beta_{\pm}$ real. The eigenvalues
$\Lambda(u)$ of the transfer matrix (\ref{transfer}) are given by
\cite{MN1}
\be
\Lambda(u) = h(u) {Q(u + p\eta)\over Q(u)} 
+ h(-u - \eta) {Q(u -p\eta)\over Q(u)}  \,,
\label{LambdaI} 
\ee
where \footnote{We find that the function $h(u)$ given by Eq.  (12) of the Addendum
\cite{MN1}, to which we now refer as $h_{old}(u)$, leads to $p-1$ ``Bethe
roots'' which actually are common to all the eigenvalues, and which therefore
should be incorporated into a new $h(u)$.  In this way, we arrive at
the expression (\ref{hfunction}), which is equal to
$h_{old}(u){\sinh(u + \eta)\over \sinh(u - \eta)}$; and at the 
$M$ value in (\ref{QI}), which is equal to $M_{old}-(p-1)$, where $M_{old}$ is
given by Eq.  (13) of the Addendum \cite{MN1}.}
\be
h(u) &=&  4\sinh^{2N+1}(u+\eta){\sinh(2u+2\eta)\over \sinh(2u+\eta)}
\sinh(u -\eta)   \non \\
&\times& \left( \cosh u +  i \sinh \beta_{-} \right) 
\left( \cosh u -  i \sinh \beta_{+} \right) \,, 
\label{hfunction}
\ee 
and
\be
Q(u) = \prod_{j=1}^{M} 
\sinh \left( {1\over 2}(u - u_{j}) \right)
\sinh \left( {1\over 2}(u + u_{j} + \eta) \right) \,,\qquad
M = N + p \,.
\label{QI}
\ee 
The zeros $u_{j}$ of $Q(u)$ satisfy the Bethe Ansatz equations
\be
{h(u_{j})\over h(-u_{j}-\eta)} = 
-{Q(u_{j}-p\eta)\over Q(u_{j}+p\eta)} \,, 
\qquad j = 1 \,, \ldots \,, M \,.
\label{BAEI1}
\ee
More explicitly, in terms of the ``shifted'' Bethe roots 
$\tilde u_{j} \equiv u_{j} + {\eta\over2}$,
\be
\lefteqn{\left({\sinh(\tilde u_{j} + {\eta\over2})\over 
       \sinh(\tilde u_{j} - {\eta\over2})}\right)^{2N+1}
{\sinh(2\tilde u_{j} + \eta)\over 
 \sinh(2\tilde u_{j} - \eta)}
{\sinh(\tilde u_{j} - {3\eta\over2})\over 
 \sinh(\tilde u_{j} + {3\eta\over2})}} \non \\
& & \times  
\left({\cosh(\tilde u_{j} - {\eta\over2}) + i\sinh \beta_{-}\over
       \cosh(\tilde u_{j} + {\eta\over2}) + i\sinh \beta_{-}}\right)
\left({\cosh(\tilde u_{j} - {\eta\over2}) - i\sinh \beta_{+}\over
       \cosh(\tilde u_{j} + {\eta\over2}) - i\sinh \beta_{+}}\right) \non \\
& & = - \prod_{k=1}^{M} 
{\cosh({1\over 2}(\tilde u_{j} - \tilde u_{k} + \eta)) \over 
 \cosh({1\over 2}(\tilde u_{j} - \tilde u_{k} - \eta))}
{\cosh({1\over 2}(\tilde u_{j} + \tilde u_{k} + \eta)) \over 
 \cosh({1\over 2}(\tilde u_{j} + \tilde u_{k} - \eta))} \,, 
\quad  j = 1 \,, \cdots \,, M \,.
\label{BAEI2}
\ee 

The energy eigenvalues are given by (\ref{Htrelation})
\be
E &=& c_{1}\Lambda'(0) + c_{2} \non \\
&=& c_{1} h(0) {Q(p \eta)\over Q(0)}\left[ {h'(0)\over h(0)} + {Q'(p \eta)\over Q(p \eta)}
-{Q'(0)\over Q(0)} \right] + c_{2} \,.
\ee
Using the fact 
\be
\Lambda(0) = h(0) {Q(p \eta)\over Q(0)} = c_{0} \,,
\ee 
where the second equality follows from (\ref{initial}), we arrive at
the result
\be
E={1\over 2} \sinh^{2}\eta \sum_{j=1}^{M}{1\over 
\sinh (\tilde u_{j} - {\eta\over2})
\sinh (\tilde u_{j} + {\eta\over2})} + {1\over 2}(N-1) \cosh \eta \,.
\label{energyI1}
\ee 

Numerical investigation of the ground state for small values of $N$
and $p$ (along the lines of \cite{NR}) suggests making a further shift
of the Bethe roots,
\be
\tilde{\tilde u}_{j} \equiv \tilde u_{j} - {i \pi\over 2} = u_{j} +
{\eta\over 2}  - {i \pi\over 2} \,,
\ee
in terms of which the Bethe Ansatz Eqs. (\ref{BAEI2}) become
\be
\lefteqn{\left({\cosh(\tilde {\tilde u}_{j} + {\eta\over2})\over 
                \cosh(\tilde {\tilde u}_{j} - {\eta\over2})}\right)^{2N+2}
{\sinh(\tilde {\tilde u}_{j} + {\eta\over2})\over 
 \sinh(\tilde {\tilde u}_{j} - {\eta\over2})}
{\cosh(\tilde {\tilde u}_{j} - {3\eta\over2})\over 
\cosh(\tilde {\tilde u}_{j} + {3\eta\over2})}} \non \\
& & \times  
{\sinh({1\over 2}(\tilde {\tilde u}_{j} + \beta_{-} - {\eta\over2}))
 \cosh({1\over 2}(\tilde {\tilde u}_{j} - \beta_{-} - {\eta\over2}))\over
 \sinh({1\over 2}(\tilde {\tilde u}_{j} + \beta_{-} + {\eta\over2}))
 \cosh({1\over 2}(\tilde {\tilde u}_{j} - \beta_{-} + {\eta\over2}))}
{\sinh({1\over 2}(\tilde {\tilde u}_{j} - \beta_{+} - {\eta\over2}))
 \cosh({1\over 2}(\tilde {\tilde u}_{j} + \beta_{+} - {\eta\over2}))\over
 \sinh({1\over 2}(\tilde {\tilde u}_{j} - \beta_{+} + {\eta\over2}))
 \cosh({1\over 2}(\tilde {\tilde u}_{j} + \beta_{+} + {\eta\over2}))} \non \\
& & = - \prod_{k=1}^{M} 
{\cosh({1\over 2}(\tilde {\tilde u}_{j} - \tilde {\tilde u}_{k} + \eta)) \over 
 \cosh({1\over 2}(\tilde {\tilde u}_{j} - \tilde {\tilde u}_{k} - \eta))}
{\sinh({1\over 2}(\tilde {\tilde u}_{j} + \tilde {\tilde u}_{k} + \eta)) \over 
 \sinh({1\over 2}(\tilde {\tilde u}_{j} + \tilde {\tilde u}_{k} - \eta))} \,, 
\quad  j = 1 \,, \cdots \,, M \,.
\label{BAEI3}
\ee 
Moreover, we find that for suitable values of the boundary parameters $\beta_{\pm}$ 
(which we discuss after Eq. (\ref{boundenergyeachI}) below), the $N+p$ Bethe roots 
$\{ \tilde{\tilde u}_{1}\,, \ldots \,, \tilde{\tilde u}_{N+p} \}$ 
for the ground state have the approximate form \footnote{Due to the periodicity
and crossing properties $Q(u+2i\pi)=Q(-u-\eta)=Q(u)$, the zeros $u_{j}$ 
are defined up to $u_{j}\mapsto u_{j}+2i\pi$ and $u_{j}\mapsto -u_{j}-\eta$,
which corresponds to $\tilde{\tilde u}_{j} \mapsto \tilde{\tilde u}_{j}
+ 2i\pi$ and $\tilde{\tilde u}_{j} \mapsto -\tilde{\tilde u}_{j}-i\pi$, 
respectively. We use these symmetries to restrict the roots to
the fundamental region $\Re e\ \tilde{\tilde u}_{j} \ge 0$
and $-\pi < \Im m\ \tilde{\tilde u}_{j} \le \pi$.}
\be
\left\{ \begin{array}{c@{\quad : \quad} l}
v_{j} \pm {i \pi\over 2} & j = 1\,, 2\,, \ldots \,, {N\over 2} \\
v_{j}^{(1)} + i \pi\,, \quad v_{j}^{(2)} & j = 1\,, 2\,, \ldots \,, {p\over 2}
\end{array} \right. \,,
\label{stringhypothesisI}
\ee 
where $\{ v_{j}\,, v_{j}^{(a)} \}$ are all {\em real} and
positive.  That is, the ground state is described by ${N\over 2}$
``strings'' of length 2, and ${p\over 2}$ pairs of strings
of length 1.

We make the ``string hypothesis'' that (\ref{stringhypothesisI})
is exactly true in the thermodynamic limit ($N \rightarrow \infty$ with
$p$ fixed).  The number of strings of length 2 therefore becomes
infinite (there is a ``sea'' of such 2-strings); and the distribution
of their centers $\{ v_{j} \}$ is described by a density function, which can
be computed from the counting function.  To this end, we form the
product of the Bethe Ansatz Eqs.  (\ref{BAEI3}) for the sea roots
$v_{j} \pm {i \pi\over 2}$. The result is given by
\be
\lefteqn{e_{1}(\lambda_{j})^{4N+4} g_{1}(\lambda_{j})^{2} 
\left[ e_{3}(\lambda_{j})^{2}
g_{1+i2b_{-}}(\lambda_{j}) g_{1-i2b_{-}}(\lambda_{j})
g_{1+i2b_{+}}(\lambda_{j}) g_{1-i2b_{+}}(\lambda_{j}) \right]^{-1}} 
\label{BAEIsea}\\
& & = \left[\prod_{k=1}^{N/2} e_{2}(\lambda_{j}-\lambda_{k}) 
e_{2}(\lambda_{j}+ \lambda_{k})\right]^{2} 
\prod_{a=1}^{2} \prod_{k=1}^{p/2} \left[ g_{2}(\lambda_{j}-\lambda_{k}^{(a)}) 
g_{2}(\lambda_{j}+\lambda_{k}^{(a)}) \right] \,, \quad j=1\,, \ldots 
\,, {N\over 2}\,, \non 
\ee 
where we have used the notation (\ref{notation}), as well as
\be
v_{j} = \mu \lambda_{j} \,, \qquad v_{j}^{(a)} = \mu 
\lambda_{j}^{(a)} \,, 
\ee
and (see \cite{AN} and references therein)
\be
e_{n}(\lambda) =
{\sinh \left(\mu  (\lambda + {i n\over 2}) \right) 
\over \sinh \left( \mu (\lambda - {i n\over 2}) \right) } \,, \qquad
g_{n}(\lambda) = e_{n}(\lambda \pm {i \pi \over 2 \mu})
= {\cosh \left(\mu  (\lambda + {i n\over 2}) \right) 
\over \cosh \left( \mu (\lambda - {i n\over 2}) \right) } \,.
\ee
Taking the logarithm of (\ref{BAEIsea}), we obtain the ground-state counting function
\be
\lefteqn{\h(\lambda) = {1\over 4 \pi}\Big\{ (4N+4) q_{1}(\lambda) + 2 r_{1}(\lambda)
-2 q_{3}(\lambda)} \non \\
& & - r_{1+i2b_{-}}(\lambda) -r_{1-i2b_{-}}(\lambda)
-r_{1+i2b_{+}}(\lambda) - r_{1-i2b_{+}}(\lambda) \non \\  
& & -2\sum_{k=1}^{N/2}\left[ q_{2}(\lambda-\lambda_{k}) +
q_{2}(\lambda + \lambda_{k})\right] - \sum_{a=1}^{2}\sum_{k=1}^{p/2}
\left[r_{2}(\lambda-\lambda_{k}^{(a)}) +
r_{2}(\lambda+\lambda_{k}^{(a)}) \right] \Big\} \,,
\ee 
where $q_{n}(\lambda)$ and $r_{n}(\lambda)$ are odd functions defined
by
\be
q_{n}(\lambda) &=& \pi + i \ln e_{n}(\lambda) 
= 2 \tan^{-1}\left( \cot(n \mu/ 2) \tanh( \mu \lambda) \right)
\,, \non \\
r_{n}(\lambda) &=&  i \ln g_{n}(\lambda) \,.
\label{logfuncts}
\ee
Noting that
\be
\sum_{k=1}^{N/2}\left[ q_{2}(\lambda-\lambda_{k}) +
q_{2}(\lambda + \lambda_{k}) \right] =  
\sum_{k=-N/2}^{N/2} q_{2}(\lambda-\lambda_{k}) -q_{2}(\lambda) \,,
\label{maneuver}
\ee
where $\lambda_{-k} \equiv -\lambda_{k}$, and letting $N$ become large, 
we obtain a linear integral equation for 
the ground-state root density $\rho(\lambda)$,
\be
\rho(\lambda) &=& {1\over N} {d \h\over d\lambda} 
 = 2 a_{1}(\lambda)
 - \int_{-\infty}^{\infty} d\lambda'\ a_{2}(\lambda - \lambda')\
 \rho(\lambda') \label{rhointegraleqn} \\ 
 &+& {1\over 2N} \Big\{ 4a_{1}(\lambda) + 2b_{1}(\lambda) - 2a_{3}(\lambda)
 + 2a_{2}(\lambda) - b_{1+2ib_{-}}(\lambda) -  b_{1-2ib_{-}}(\lambda) \non \\
 &-&  b_{1+2ib_{+}}(\lambda) -  b_{1-2ib_{+}}(\lambda) 
 - \sum_{a=1}^{2}\sum_{k=1}^{p/2}
\left[b_{2}(\lambda-\lambda_{k}^{(a)}) +
b_{2}(\lambda+\lambda_{k}^{(a)}) \right]\Big\} \,, 
\ee
where we have ignored corrections of higher order in $1/N$ when
passing from a sum to an integral, and we have introduced the
notations
\be
a_n(\lambda) &=& {1\over 2\pi} {d \over d\lambda} q_n (\lambda)
= {\mu \over \pi} 
{\sin (n \mu)\over \cosh(2 \mu \lambda) - \cos (n \mu)} \,, \non \\
b_n(\lambda) &=& {1\over 2\pi} {d \over d\lambda} r_n (\lambda)
= -{\mu \over \pi} 
{\sin (n \mu)\over \cosh(2 \mu \lambda) + \cos (n \mu)} \,. 
\label{anbn}
\ee 
Using Fourier transforms, we obtain \footnote{Our 
conventions are
\be
\hat f(\omega) \equiv \int_{-\infty}^\infty e^{i \omega \lambda}\ 
f(\lambda)\ d\lambda \,, \qquad\qquad
f(\lambda) = {1\over 2\pi} \int_{-\infty}^\infty e^{-i \omega \lambda}\ 
\hat f(\omega)\ d\omega \,. \non 
\ee} 
\be
\rho(\lambda) = 2 s(\lambda) + {1\over N} R(\lambda) \,,
\label{rho}
\ee
where
\be
s(\lambda) = {1\over 2\pi} \int_{-\infty}^{\infty} d\omega\ 
e^{-i \omega \lambda} {1\over 2 \cosh(\omega/2)} 
= {1\over 2 \cosh (\pi \lambda)} \,,
\ee
and 
\be
\hat R(\omega) &=& {1\over 2 \left(1+\hat a_{2}(\omega) \right)} 
\Big\{ 4 \hat a_{1}(\omega) + 2 \hat b_{1}(\omega) - 2 \hat a_{3}(\omega)
 + 2 \hat a_{2}(\omega) - \hat b_{1+2ib_{-}}(\omega) -  \hat b_{1-2ib_{-}}(\omega) \non \\
 &-&  \hat b_{1+2ib_{+}}(\omega) -  \hat b_{1-2ib_{+}}(\omega) 
 - 2\sum_{a=1}^{2}\sum_{k=1}^{p/2}
\cos (\omega \lambda_{k}^{(a)})\ \hat b_{2}(\omega)  \Big\} \,,
\ee
with
\be
\hat a_{n}(\omega) &=& \sgn(n) {\sinh \left( (\nu  - |n|) 
\omega / 2 \right) \over
\sinh \left( \nu \omega / 2 \right)} \,,
\qquad 0 \le |n| < 2 \nu  \,, \label{fourier1} \\
\hat b_{n}(\omega) &=&
-{\sinh \left( n \omega / 2 \right) \over
\sinh \left( \nu \omega / 2 \right)} \,,
\qquad \qquad\qquad\quad  0 < \Re e\ n < \nu  \,.
\label{fourier2}
\ee

Turning now to the expression (\ref{energyI1}) for the energy, and
invoking again the string hypothesis (\ref{stringhypothesisI}),
we see that
\be
E &=& -{1\over 2} \sinh^{2}\eta \sum_{j=1}^{M}{1\over 
\cosh (\tilde{\tilde u}_{j} - {\eta\over2})
\cosh (\tilde{\tilde u}_{j} + {\eta\over2})} + {1\over 2}(N-1) \cosh 
\eta \non \\
 &=& -{1\over 2} \sinh^{2}\eta \Big\{
-2\sum_{j=1}^{N/2}{1\over 
\sinh (v_{j} - {\eta\over2})
\sinh (v_{j} + {\eta\over2})} +\sum_{a=1}^{2}\sum_{j=1}^{p/2}{1\over 
\cosh (v_{j}^{(a)} - {\eta\over2})
\cosh (v_{j}^{(a)} + {\eta\over2})} \Big\} \non \\
& & + {1\over 2}(N-1) \cosh \eta \non \\
&=& - {2\pi \sin \mu\over \mu} \Big\{ 
\sum_{j=1}^{N/2} a_{1}(\lambda_{j}) 
+ {1\over 2} \sum_{a=1}^{2}\sum_{j=1}^{p/2} b_{1}(\lambda_{j}^{(a)} ) \Big\}
+ {1\over 2}(N-1) \cos \mu \,.
\label{energyI2}
\ee 
Repeating the maneuver (\ref{maneuver}) in the summation over the
centers of the sea roots, and letting $N$ become large, 
we obtain
\be
E &=& - {\pi \sin \mu\over \mu} \Big\{ 
\sum_{j=-N/2}^{N/2} a_{1}(\lambda_{j}) - a_{1}(0)
+  \sum_{a=1}^{2}\sum_{j=1}^{p/2} b_{1}(\lambda_{j}^{(a)} ) \Big\}
+ {1\over 2}(N-1) \cos \mu   \\
&=& - {\pi \sin \mu\over \mu} \Big\{ 
N\int_{-\infty}^{\infty}d\lambda\ a_{1}(\lambda)\ \rho(\lambda) - a_{1}(0) 
+  \sum_{a=1}^{2}\sum_{j=1}^{p/2} b_{1}(\lambda_{j}^{(a)} ) \Big\}
+ {1\over 2}(N-1) \cos \mu \non \,,
\label{energyI3}
\ee 
where again we ignore corrections that are higher order in $1/N$.
Substituting the result (\ref{rho}) for the root density, we obtain
\be
E = E_{bulk} + E_{boundary} \,,
\ee
where the bulk (order $N$) energy is given by
\be
E_{bulk} &=& - {2N \pi \sin \mu\over \mu} 
\int_{-\infty}^{\infty}d\lambda\ a_{1}(\lambda)\ s(\lambda) 
+ {1\over 2}N \cos \mu \non \\
&=&  - N \sin^{2} \mu \int_{-\infty}^{\infty}
d\lambda\ {1\over \left[\cosh(2 \mu \lambda) - \cos \mu \right] 
\cosh (\pi \lambda)} +  {1\over 2}N \cos \mu \,,
\label{bulkenergy}
\ee
which agrees with the well-known result \cite{YY}.  Moreover, the
boundary (order $1$) energy is given by
\be
E_{boundary} = - {\pi \sin \mu\over \mu} \Big\{
I +  \sum_{a=1}^{2}\sum_{j=1}^{p/2} b_{1}(\lambda_{j}^{(a)} ) \Big\}
 -{1\over 2}\cos \mu \,,
\label{boundenergyI}
\ee
where $I$ is the integral 
\be
I &=& \int_{-\infty}^{\infty}d\lambda\ a_{1}(\lambda) \left[
R(\lambda) - \delta(\lambda) \right] 
= {1\over 2\pi}  \int_{-\infty}^{\infty} d\omega\ \hat a_{1}(\omega)
\left[ \hat R(\omega) - 1 \right] \non \\
&=& -\sum_{a=1}^{2}\sum_{j=1}^{p/2} b_{1}(\lambda_{j}^{(a)} )
+ {1\over 2\pi}  \int_{-\infty}^{\infty} d\omega\ \hat s(\omega)
\Big\{ 2 \hat a_{1}(\omega) + \hat b_{1}(\omega) -  \hat a_{3}(\omega) - 
1 \non \\
& & - {1\over 2}\left[ \hat b_{1+2ib_{-}}(\omega) + \hat b_{1-2ib_{-}}(\omega) 
 + \hat b_{1+2ib_{+}}(\omega) +  \hat b_{1-2ib_{+}}(\omega) \right] 
 \Big\} \,, \label{integralI}
\ee
where we have used the fact $\hat s(\omega) \hat b_{2}(\omega) = \hat
b_{1}(\omega)$. 
Remarkably, the $\lambda_{j}^{(a)}$-dependent contribution in
(\ref{boundenergyI}) is exactly canceled by an opposite contribution
from the integral $I$ (\ref{integralI}).  Writing the boundary energy
as the sum of contributions from the left and right boundaries,
$E_{boundary}= E_{boundary}^{-} + E_{boundary}^{+}$, we conclude that
the energy contribution from each boundary is given by
\be
E_{boundary}^{\pm}&=& - {\sin \mu\over 2\mu} 
\int_{-\infty}^{\infty} d\omega\ 
{1\over 2\cosh (\omega/ 2)}
\Big\{ 
{\sinh((\nu-2)\omega/4) \over 2\sinh(\nu \omega/4)} 
-{1\over 2} \non \\
&+& {\sinh(\omega/2) \cosh((\nu-2)\omega/2)\over 
\sinh(\nu\omega/2)} + 
{\sinh(\omega/2) \cos (b_{\pm}\omega) \over 
\sinh(\nu \omega/2)} \Big\} -{1\over 4}\cos \mu \,.
\label{boundenergyeachI}
\ee
One can verify that this result coincides with the result
(\ref{ANboundenergy}) with $a_{\pm}=1$.  As shown in the Appendix,
the integrals in (\ref{bulkenergy}) and (\ref{boundenergyeachI}) 
(with $p$ even) can be evaluated analytically.

We have derived the result (\ref{boundenergyeachI}) for the boundary
energy under the assumption that the Bethe roots for the ground state
have the form (\ref{stringhypothesisI}), which is true only for
suitable values of the boundary parameters $\beta_{\pm}$.  For
example, the shaded region in Fig. \ref{fig:p=4,n=2} denotes 
the region of parameter
space for which the ground-state Bethe roots have the form
(\ref{stringhypothesisI}) for $p=4$ and $N=2$.  For parameter values
outside the shaded region, one or more of the Bethe roots has an
imaginary part which is {\em not} a multiple of $\pi/2$ and which
evidently depends on the parameter values (but in a manner which we
have not yet explicitly determined).  As $p$ increases, the
figure is similar, except that the shaded region moves further away
from the origin.

\begin{figure}[htb]
	\centering
	\includegraphics[width=0.50\textwidth]{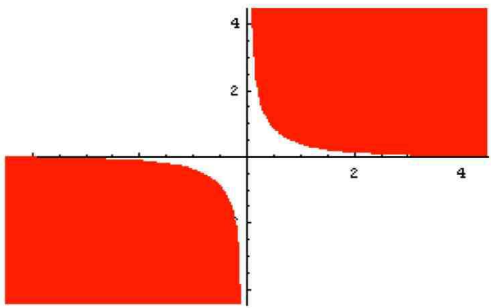}
	\caption[xxx]{\parbox[t]{0.6\textwidth}{
	Shaded region denotes region of the $(\beta_{+}\,, \beta_{-})$
	plane for which the ground-state Bethe roots have the form 
	(\ref{stringhypothesisI}) for $p=4$ and $N=2$.}
	}
	\label{fig:p=4,n=2}
\end{figure}

\begin{figure}[htb]
	\centering
	\includegraphics[width=0.25\textwidth]{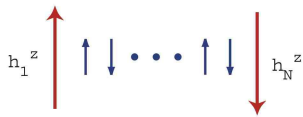}
	\caption[xxx]{\parbox[t]{0.6\textwidth}{
	Antiparallel boundary fields (big, red) {\bf are} compatible with antiferromagnetic
	alignment of spins (small, blue)}
	}
	\label{fig:antiparallel}
\end{figure}

A qualitative explanation of these features can be deduced from a
short heuristic argument.  Indeed, let us rewrite the Hamiltonian
(\ref{HamiltonianI}) as
\be
{\mathcal H} = {\mathcal H}_{0}
+ h^{z}_{1} \sigma_{1}^{z}
+ h^{x}_{1} \sigma_{1}^{x} 
+ h^{z}_{N} \sigma_{N}^{z}
+ h^{x}_{N} \sigma_{N}^{x}   \,, 
\ee
where the boundary magnetic fields are given by
\be
h^{z}_{1} &=& {1\over 2} \ch \eta \tnh \beta_{-} \,, 
\quad h^{x}_{1} = {1\over 2} \sech \beta_{-} \non \\ 
h^{z}_{N} &=& -{1\over 2} \ch \eta \tnh \beta_{+} \,, 
\quad h^{x}_{N} = {1\over 2} \sech \beta_{+} \,.
\ee
For $\beta_{+} \beta_{-} >\!\!> 0$ (i.e., the shaded regions in Fig. 1), the
boundary fields in the $x$ direction are small; moreover, $h^{z}_{1}
h^{z}_{N} < 0$; i.e., the boundary fields in the $z$ direction have
antiparallel orientations, which (since $N$ is even) is compatible with
a N\'eel-like (antiferromagnetic) alignment of the spins. (See Fig. 
\ref{fig:antiparallel}.) Hence, the ground state and corresponding Bethe 
roots are ``simple''. 

\begin{figure}[tb]
	\centering
	\includegraphics[width=0.25\textwidth]{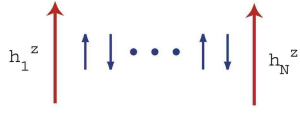}
	\caption[xxx]{\parbox[t]{0.6\textwidth}{
	Parallel boundary fields (big, red) are {\bf not} compatible with antiferromagnetic
	alignment of spins (small, blue)}
	}
	\label{fig:parallel}
\end{figure}

On the other hand, if $|\beta_{\pm}|$ are small (the unshaded region
near the origin of Fig.  1), then the boundary fields in the $x$
direction are large.  Also, if $\beta_{+} \beta_{-} < 0$ (the second
and fourth quadrants of Fig.1, which are also unshaded), then
$h^{z}_{1} h^{z}_{N} > 0$; i.e., the boundary fields in the $z$
direction are parallel, which can lead to
``frustration''.  (See Fig.  \ref{fig:parallel}.)  For these cases, the ground states
and corresponding Bethe roots are ``complicated''.

\section{Case II: $p$ odd}\label{sec:caseII}

For Case II (\ref{caseII}), the Hamiltonian (\ref{Hamiltonian}) becomes
\be
{\mathcal H} = {\mathcal H}_{0}
+ {1\over 2} \sh \eta \left( 
\csch \alpha_{-} \sigma_{1}^{x} 
+ \csch \alpha_{+} \sigma_{N}^{x} \right)  \,.
\label{HamiltonianII} 
\ee
We restrict $\alpha_{\pm}$ to be purely imaginary in order for the
Hamiltonian to be Hermitian. We use the periodicity 
$\alpha_{\pm} \mapsto \alpha_{\pm} + 2\pi i$
of the transfer matrix to further restrict $\alpha_{\pm}$ to the 
fundamental domain $-\pi \le \Im m\ \alpha_{\pm} < \pi$.
The eigenvalues
$\Lambda(u)$ of the transfer matrix (\ref{transfer}) are given by
\cite{MN2} \footnote{The function $Q_{2}(u)$ here as well as its
zeros $\{ u_{j}^{(2)} \}$ are shifted by $\eta$ with  
respect to the corresponding quantities in \cite{MN2}, to which we now refer 
as ``old''; i.e., $Q_{2}(u) = Q_{2}^{old}(u - \eta)$ and
$u_{j}^{(2)} = u_{j}^{(2)\ old}- \eta$.}
\be
\Lambda(u) &=&
 {\delta(u-\eta)\over h^{(2)}(u-\eta)} {Q_{2}(u-\eta)\over 
Q_{1}(u)}
+ {\delta(u)\over h^{(1)}(u)} {Q_{2}(u+\eta)\over Q_{1}(u)} 
\,, \non  \\
 &=& 
h^{(1)}(u-\eta) {Q_{1}(u-\eta)\over Q_{2}(u)}
 +  h^{(2)}(u) {Q_{1}(u+\eta)\over 
Q_{2}(u)}
 \,,
\label{TQ}
\ee
where \footnote{Similarly to Case I, we find that the functions
$h^{(1)}(u)$ given in Eqs.  (A.5), (A.6) of \cite{MN2} lead to ``Bethe
roots'' which actually are common to all the eigenvalues, and which
therefore should be incorporated into a new $h^{(1)}(u)$.  In this
way, we arrive at the expression for $h^{(1)}(u)$ in
(\ref{h1function}) and the corresponding $M_{a}$ values in
(\ref{MII}).}
\be
h^{(1)}(u) &=& {8\sinh^{2N+1}(u+2\eta)\cosh^{2}(u+\eta) \cosh(u+2\eta)\over 
\sinh(2u+3\eta)} \,, \quad h^{(2)}(u) = h^{(1)}(-u-2\eta) \,, \label{h1function} \\
\delta(u) &=& h^{(1)}(u) h^{(2)}(u) 
\sinh(u+\eta+\alpha_{-})\sinh(u+\eta-\alpha_{-})
\sinh(u+\eta+\alpha_{+})\sinh(u+\eta-\alpha_{+}) \,, \non
\ee 
and
\be
Q_{a}(u) = \prod_{j=1}^{M_{a}} 
\sinh (u - u_{j}^{(a)}) \sinh (u + u_{j}^{(a)} + \eta) \,, \qquad a = 
1\,, 2\,, 
\label{QII}
\ee
with
\be
M_{1} = {1\over 2}(N+p+1) \,, \qquad M_{2} = {1\over 2}(N+p-1) \,.
\label{MII}
\ee 
As remarked in the Introduction, the expressions for the eigenvalues 
(\ref{TQ}) correspond to generalized $T-Q$ relations 
(\ref{generalizedTQ}). For generic values of $\alpha_{\pm}$, we have
not managed to reformulate this solution in terms of a single $Q(u)$.
The  zeros $\{ u_{j}^{(a)} \}$ of $Q_{a}(u)$ are given by the
Bethe Ansatz equations,
\be
{\delta(u_{j}^{(1)})\ h^{(2)}(u_{j}^{(1)}-\eta)
\over \delta(u_{j}^{(1)}-\eta)\ h^{(1)}(u_{j}^{(1)})} 
&=&-{Q_{2}(u_{j}^{(1)}-\eta)\over Q_{2}(u_{j}^{(1)}+\eta)} \,, \qquad j =
1\,, 2\,, \ldots \,, M_{1} \,, \non \\
{h^{(1)}(u_{j}^{(2)}-\eta)\over h^{(2)}(u_{j}^{(2)})}
&=&-{Q_{1}(u_{j}^{(2)}+\eta)\over Q_{1}(u_{j}^{(2)}-\eta)} \,, \qquad j =
1\,, 2\,, \ldots \,, M_{2} \,.
\label{BAEII1}
\ee 
In terms of the ``shifted'' Bethe roots 
$\tilde u_{j}^{(a)} \equiv u_{j}^{(a)} + {\eta\over2}$, the Bethe Ansatz
equations are 
\be
\lefteqn{\left({\sinh(\tilde u_{j}^{(1)} + {\eta\over2})\over 
       \sinh(\tilde u_{j}^{(1)} - {\eta\over2})}\right)^{2N+1}
{\cosh(\tilde u_{j}^{(1)} - {\eta\over2})\over 
 \cosh(\tilde u_{j}^{(1)} + {\eta\over2})}} \non \\
& & \times  
{\sinh(\tilde u_{j}^{(1)} +\alpha_{-} - {\eta\over2})\over 
 \sinh(\tilde u_{j}^{(1)} -\alpha_{-} + {\eta\over2})}
{\sinh(\tilde u_{j}^{(1)} -\alpha_{-} - {\eta\over2})\over 
 \sinh(\tilde u_{j}^{(1)} +\alpha_{-} + {\eta\over2})}
{\sinh(\tilde u_{j}^{(1)} +\alpha_{+} - {\eta\over2})\over 
 \sinh(\tilde u_{j}^{(1)} -\alpha_{+} + {\eta\over2})}
{\sinh(\tilde u_{j}^{(1)} -\alpha_{+} - {\eta\over2})\over 
 \sinh(\tilde u_{j}^{(1)} +\alpha_{+} + {\eta\over2})}
 \non \\
& & = - \prod_{k=1}^{M_{2}} 
{\sinh(\tilde u_{j}^{(1)} - \tilde u_{k}^{(2)} + \eta) \over 
 \sinh(\tilde u_{j}^{(1)} - \tilde u_{k}^{(2)} - \eta)}
{\sinh(\tilde u_{j}^{(1)} + \tilde u_{k}^{(2)} + \eta) \over 
 \sinh(\tilde u_{j}^{(1)} + \tilde u_{k}^{(2)} - \eta)}
\,, 
\quad  j = 1 \,, \cdots \,, M_{1} \,.
\label{BAEII2a}
\ee 
and
\be
\lefteqn{\left({\sinh(\tilde u_{j}^{(2)} + {\eta\over2})\over 
                \sinh(\tilde u_{j}^{(2)} - {\eta\over2})}\right)^{2N+1}
{\cosh(\tilde u_{j}^{(2)} - {\eta\over2})\over 
 \cosh(\tilde u_{j}^{(2)} + {\eta\over2})}} \non \\
& & = - \prod_{k=1}^{M_{1}} 
{\sinh(\tilde u_{j}^{(2)} - \tilde u_{k}^{(1)} + \eta) \over 
 \sinh(\tilde u_{j}^{(2)} - \tilde u_{k}^{(1)} - \eta)}
{\sinh(\tilde u_{j}^{(2)} + \tilde u_{k}^{(1)} + \eta) \over 
 \sinh(\tilde u_{j}^{(2)} + \tilde u_{k}^{(1)} - \eta)}
\,, 
\quad  j = 1 \,, \cdots \,, M_{2} \,.
\label{BAEII2b}
\ee 

The energy is given by
\be
E={1\over 2} \sinh^{2}\eta 
\sum_{a=1}^{2}\sum_{j=1}^{M_{a}}{1\over 
\sinh (\tilde u_{j}^{(a)} - {\eta\over2})
\sinh (\tilde u_{j}^{(a)} + {\eta\over2})} + {1\over 2}(N-1) \cosh 
\eta \,.
\label{energyII1}
\ee 
Indeed, for $p>1$, we obtain this result by following steps similar to
those leading to (\ref{energyI1}).  For $p=1$, we use
(\ref{initialp1}) and (\ref{Htrelationp1}) instead of (\ref{initial})
and (\ref{Htrelation}); nevertheless, the result (\ref{energyII1})
holds also for $p=1$.

From numerical studies for small values of $N$
and $p$, and for suitable values of the boundary parameters $\alpha_{\pm}$ 
(which we discuss after Eq. (\ref{boundenergyeachII}) below), we find that the ground state is 
described by Bethe roots $\{ \tilde u_{j}^{(1)} \}$ and $\{ \tilde
u_{j}^{(2)} \}$
of the form \footnote{The periodicity
and crossing properties of $Q_{a}(u)$ imply that the zeros $u_{j}^{(a)}$ 
are defined up to $u_{j}^{(a)} \mapsto u_{j}^{(a)}+i\pi$ and 
$u_{j}^{(a)}\mapsto -u_{j}^{(a)}-\eta$,
which corresponds to $\tilde u_{j}^{(a)} \mapsto \tilde u_{j}^{(a)}
+ i\pi$ and $\tilde u_{j}^{(a)} \mapsto -\tilde u_{j}^{(a)}$, 
respectively. We use these symmetries to restrict the roots to
the fundamental region $\Re e\ \tilde u_{j}^{(a)} \ge 0$
and $-{\pi\over 2} < \Im m\ \tilde u_{j}^{(a)} \le {\pi\over 2}$.}
\be
\left\{ \begin{array}{c@{\quad : \quad} l}
v_{j}^{(1,1)}                      & j = 1\,, 2\,, \ldots \,, {N\over 2} \\
v_{j}^{(1,2)} + {i \pi\over 2} \,, & j = 1\,, 2\,, \ldots \,, {p+1\over 2}
\end{array} \right. \,, \qquad
\left\{ \begin{array}{c@{\quad : \quad} l}
v_{j}^{(2,1)}                      & j = 1\,, 2\,, \ldots \,, {N\over 2} \\
v_{j}^{(2,2)} + {i \pi\over 2} \,, & j = 1\,, 2\,, \ldots \,, {p-1\over 2}
\end{array} \right. \,,
\label{stringhypothesisII}
\ee 
respectively, where $\{ v_{j}^{(a,b)} \}$ are  all real and positive.

We make the ``string hypothesis'' that (\ref{stringhypothesisII})
remains true in the thermodynamic limit ($N \rightarrow \infty$ with
$p$ fixed). That is, that the Bethe roots $\{ \tilde u_{j}^{(a)} \}$ 
for the ground state have the form
\be
\left\{ \begin{array}{c@{\quad : \quad} l}
v_{j}^{(a,1)}                      & j = 1\,, 2\,, \ldots \,, M_{(a,1)} \\
v_{j}^{(a,2)} + {i \pi\over 2} \,, & j = 1\,, 2\,, \ldots \,, M_{(a,2)}
\end{array} \right. \,, \qquad a = 1\,, 2 \,, 
\label{stringhypothesisIIc}
\ee 
where $\{ v_{j}^{(a,b)} \}$ are  all real and
positive; also, $M_{(1,1)}= M_{(2,1)} = {N\over 2}$, and $M_{(1,2)}={p+1\over 
2}$, $M_{(2,2)}={p-1\over 2}$. Evidently there are two ``seas'' of real roots,
namely $\{ v_{j}^{(1,1)} \}$ and $\{ v_{j}^{(2,1)} \}$.

We now proceed to compute the boundary energy, using 
notations similar to those in Case I. Defining
\be
v_{j}^{(a,b)} = \mu \lambda_{j}^{(a,b)}  \,, 
\ee
the Bethe Ansatz equations (\ref{BAEII2a}), (\ref{BAEII2b}) for the sea roots are
\be
\lefteqn{e_{1}(\lambda_{j}^{(1,1)})^{2N+1} 
\left[ g_{1}(\lambda_{j}^{(1,1)}) 
e_{1+2a_{-}}(\lambda_{j}^{(1,1)}) e_{1-2a_{-}}(\lambda_{j}^{(1,1)})
e_{1+2a_{+}}(\lambda_{j}^{(1,1)}) e_{1-2a_{+}}(\lambda_{j}^{(1,1)}) \right]^{-1}} 
\label{BAEIIsea1} \\
& & = -\prod_{k=1}^{N/2} \left[
e_{2}(\lambda_{j}^{(1,1)} - \lambda_{k}^{(2,1)}) 
e_{2}(\lambda_{j}^{(1,1)} + \lambda_{k}^{(2,1)})\right] 
 \prod_{k=1}^{(p-1)/2} \left[ 
g_{2}(\lambda_{j}^{(1,1)} - \lambda_{k}^{(2,2)}) 
g_{2}(\lambda_{j}^{(1,1)} + \lambda_{k}^{(2,2)}) \right] \,, \non 
\ee 
and
\be
\lefteqn{e_{1}(\lambda_{j}^{(2,1)})^{2N+1} 
g_{1}(\lambda_{j}^{(2,1)})^{-1}} 
\label{BAEIIsea2} \\
& & = -\prod_{k=1}^{N/2} \left[
e_{2}(\lambda_{j}^{(2,1)} - \lambda_{k}^{(1,1)}) 
e_{2}(\lambda_{j}^{(2,1)} + \lambda_{k}^{(1,1)})\right] 
\prod_{k=1}^{(p+1)/2} \left[  
g_{2}(\lambda_{j}^{(2,1)} - \lambda_{k}^{(1,2)}) 
g_{2}(\lambda_{j}^{(2,1)} + \lambda_{k}^{(1,2)}) \right] \,, \non 
\ee 
respectively, where $j=1\,, \ldots \,, {N\over 2}$. The corresponding  
ground-state counting functions are 
\be
\lefteqn{\h^{(1)}(\lambda) = {1\over 2 \pi}\Big\{ (2N+1) 
q_{1}(\lambda) - r_{1}(\lambda)
- q_{1+2a_{-}}(\lambda) - q_{1-2a_{-}}(\lambda)
- q_{1+2a_{+}}(\lambda) - q_{1-2a_{+}}(\lambda)} \non \\  
& & -\sum_{k=1}^{N/2}\left[ 
q_{2}(\lambda - \lambda_{k}^{(2,1)}) +
q_{2}(\lambda + \lambda_{k}^{(2,1)})\right] 
- \sum_{k=1}^{(p-1)/2} \left[
r_{2}(\lambda - \lambda_{k}^{(2,2)}) +
r_{2}(\lambda + \lambda_{k}^{(2,2)}) \right] \Big\} \,,
\ee 
and
\be
\lefteqn{\h^{(2)}(\lambda) = {1\over 2 \pi}\Big\{ (2N+1) 
q_{1}(\lambda) - r_{1}(\lambda)} \non \\
& & -\sum_{k=1}^{N/2}\left[ 
q_{2}(\lambda - \lambda_{k}^{(1,1)}) +
q_{2}(\lambda + \lambda_{k}^{(1,1)})\right] 
- \sum_{k=1}^{(p+1)/2} \left[
r_{2}(\lambda - \lambda_{k}^{(1,2)}) +
r_{2}(\lambda + \lambda_{k}^{(1,2)}) \right] \Big\} \,.
\ee 
Repeating the maneuver (\ref{maneuver})  in the summations over the
sea roots, and letting $N$ become large, 
we obtain a pair of coupled linear integral equations for 
the ground-state root densities $\rho^{(a)}(\lambda)$,
\be
\rho^{(1)}(\lambda) &=& {1\over N} {d \h^{(1)}\over d\lambda} 
 = 2 a_{1}(\lambda)
 - \int_{-\infty}^{\infty} d\lambda'\ a_{2}(\lambda - \lambda')\
 \rho^{(2)}(\lambda')  \non \\ 
 &+& {1\over N} \Big\{ a_{1}(\lambda) + a_{2}(\lambda)  - b_{1}(\lambda) 
    - a_{1+2a_{-}}(\lambda) -  a_{1-2a_{-}}(\lambda) \non \\
 &-&  a_{1+2a_{+}}(\lambda) -  a_{1-2a_{+}}(\lambda) 
 - \sum_{k=1}^{(p-1)/2}
\left[b_{2}(\lambda-\lambda_{k}^{(2,2)}) +
b_{2}(\lambda+\lambda_{k}^{(2,2)}) \right]\Big\} \,, 
\ee
and 
\be
\rho^{(2)}(\lambda) &=& {1\over N} {d \h^{(2)}\over d\lambda} 
 = 2 a_{1}(\lambda)
 - \int_{-\infty}^{\infty} d\lambda'\ a_{2}(\lambda - \lambda')\
 \rho^{(1)}(\lambda')  \non \\ 
 &+& {1\over N} \Big\{ a_{1}(\lambda) + a_{2}(\lambda)  - b_{1}(\lambda) 
 - \sum_{k=1}^{(p+1)/2}
\left[b_{2}(\lambda-\lambda_{k}^{(1,2)}) +
b_{2}(\lambda+\lambda_{k}^{(1,2)}) \right]\Big\} \,.
\ee
It is straightforward to solve by Fourier transforms for the 
individual root densities. However, we shall see that the energy depends only 
on the sum of the root densities, which is given by
\be
\rho^{(1)}(\lambda) + \rho^{(2)}(\lambda) = 4 s(\lambda) 
+  {1\over N} {\cal R}(\lambda) \,,
\label{rhosum}
\ee 
where
\be
\hat {\cal R}(\omega) = &=& {1\over 1+\hat a_{2}(\omega)} 
\Big\{ 2\hat a_{1}(\omega) + 2\hat a_{2}(\omega) - 2 \hat b_{1}(\omega) 
  - \hat a_{1+2a_{-}}(\omega) -  \hat a_{1-2a_{-}}(\omega) \non \\
 &-&  \hat a_{1+2a_{+}}(\omega) -  \hat a_{1-2a_{+}}(\omega) 
 - 2\sum_{a=1}^{2}\sum_{k=1}^{M_{(a,2)}}
\cos (\omega \lambda_{k}^{(a,2)})\ \hat b_{2}(\omega)  \Big\} \,.
\ee 

The expression (\ref{energyII1}) for the energy and the string
hypothesis (\ref{stringhypothesisII}) imply
\be
E &=&  -{1\over 2} \sinh^{2}\eta \Big\{
-\sum_{a=1}^{2}\sum_{j=1}^{N/2}{1\over 
\sinh (v_{j}^{(a,1)} - {\eta\over2})
\sinh (v_{j}^{(a,1)} + {\eta\over2})} \non \\
& & -\sum_{a=1}^{2}\sum_{j=1}^{M_{(a,2)}}{1\over 
\cosh (v_{j}^{(a,2)} - {\eta\over2})
\cosh (v_{j}^{(a,2)} + {\eta\over2})} \Big\} 
 + {1\over 2}(N-1) \cosh \eta \non \\
&=& - {\pi \sin \mu\over \mu} \Big\{ 
\sum_{a=1}^{2}\sum_{j=1}^{N/2} a_{1}(\lambda_{j}^{(a,1)}) +
\sum_{a=1}^{2}\sum_{j=1}^{M_{(a,2)}} b_{1}(\lambda_{j}^{(a,2)} ) \Big\}
+ {1\over 2}(N-1) \cos \mu \non  \\
&=& - {\pi \sin \mu\over \mu} \Big\{ 
{1\over 2}\sum_{a=1}^{2}\sum_{j=-N/2}^{N/2} 
a_{1}(\lambda_{j}^{(a,1)}) -a_{1}(0)
+ \sum_{a=1}^{2}\sum_{j=1}^{M_{(a,2)}} b_{1}(\lambda_{j}^{(a,2)} ) \Big\}
+ {1\over 2}(N-1) \cos \mu \non \\
&=& - {\pi \sin \mu\over \mu} \Big\{ 
{N\over 2}\int_{-\infty}^{\infty}d\lambda\ a_{1}(\lambda)
\left[ \rho^{(1)}(\lambda) + \rho^{(2)}(\lambda) \right]- a_{1}(0) 
\non\\
& & +  \sum_{a=1}^{2}\sum_{j=1}^{M_{(a,2)}} b_{1}(\lambda_{j}^{(a,2)} ) \Big\}
+ {1\over 2}(N-1) \cos \mu  \,. \label{energyII2} 
\ee 
Substituting the result (\ref{rhosum}) for the sum of the root 
densities, we obtain
\be
E = E_{bulk} + E_{boundary} \,,
\ee
where the bulk (order $N$) energy is again given by (\ref{bulkenergy}),
and the boundary (order $1$) energy is given by
\be
E_{boundary} = - {\pi \sin \mu\over \mu} \Big\{
{\cal I} +  \sum_{a=1}^{2}\sum_{j=1}^{M_{(a,2)}} b_{1}(\lambda_{j}^{(a,2)} ) \Big\}
 -{1\over 2}\cos \mu \,,
\label{boundenergyII}
\ee
where ${\cal I}$ is the integral 
\be
{\cal I} &=& {1\over 2}\int_{-\infty}^{\infty}d\lambda\ a_{1}(\lambda) \left[
{\cal R}(\lambda) - 2\delta(\lambda) \right] 
= {1\over 4\pi}  \int_{-\infty}^{\infty} d\omega\ \hat a_{1}(\omega)
\left[ \hat {\cal R}(\omega) - 2 \right] \non \\
&=& -\sum_{a=1}^{2}\sum_{j=1}^{M_{(a,2)}} b_{1}(\lambda_{j}^{(a,2)} )
+ {1\over 2\pi}  \int_{-\infty}^{\infty} d\omega\ \hat s(\omega)
\Big\{ \hat a_{1}(\omega) - \hat b_{1}(\omega)  - 1 \non \\
& & - {1\over 2}\left[ \hat a_{1+2a_{-}}(\omega) + \hat a_{1-2a_{-}}(\omega) 
 + \hat a_{1+2a_{+}}(\omega) +  \hat a_{1-2a_{+}}(\omega) \right] 
 \Big\} \,. \label{integralII}
\ee
Once again there is a remarkable cancellation among terms involving
Bethe roots which are not parts of the seas, namely, $\lambda_{j}^{(a,2)}$.
Writing $E_{boundary}$ as the sum of contributions from the left and
right boundaries, we conclude that for the parameter values
\footnote{This restriction arises when using the Fourier
transform result (\ref{fourier1}) to evaluate $\hat a_{1+2a_{\pm}}(\omega) + \hat
a_{1-2a_{\pm}}(\omega)$.}
\be
{1\over 2} \le |a_{\pm}| < \nu -{1\over 2} \,,
\label{restriction}
\ee
the energy contribution from each
boundary is given by
\be
E_{boundary}^{\pm}&=& - {\sin \mu\over 2\mu} 
\int_{-\infty}^{\infty} d\omega\ 
{1\over 2\cosh (\omega/ 2)}
\Big\{ 
{\cosh((\nu-2)\omega/4) \over 2\cosh(\nu \omega/4)} 
-{1\over 2} \non\\
&+& {\sinh(\omega/2) \cosh ((\nu-2|a_{\pm}|)\omega/2) \over 
\sinh(\nu \omega/2)} \Big\} -{1\over 4}\cos \mu \,.
\label{boundenergyeachII}  
\ee
This result agrees with the result (\ref{ANboundenergy}) with $b_{\pm}=0$
and $a_{\pm}$ values (\ref{restriction}).  As shown in the Appendix,
the integrals (with $p$ odd) can also be evaluated analytically.

We have derived the above result for the boundary energy under the
assumption that the Bethe roots for the ground state have the form
(\ref{stringhypothesisII}), which is true only for suitable values of
the boundary parameters $a_{\pm}$, namely,
\be
{\nu-1 \over 2} < |a_{\pm}| < {\nu+1 \over 2} \,, 
\qquad a_{+} a_{-} > 0\,, 
\label{niceregionII}
\ee 
where $\nu=p+1$.  For parameter values outside the region
(\ref{niceregionII}), one or more of the Bethe roots has an imaginary
part which is {\em not} a multiple of $\pi/2$ and which evidently
depends on the parameter values. One can verify that the region
(\ref{niceregionII}) is contained in the region (\ref{restriction}).

As in Case I, it is possible to give a qualitative explanation of 
the restriction (\ref{niceregionII}) by a heuristic argument. Indeed,
let us rewrite the Hamiltonian (\ref{HamiltonianII}) as
\be
{\mathcal H} = {\mathcal H}_{0}
+ h^{x}_{1} \sigma_{1}^{x} 
+ h^{x}_{N} \sigma_{N}^{x}   \,, 
\ee
where the boundary magnetic fields are given by
\be
h^{x}_{1} = {1\over 2} \sinh \eta  \csch \alpha_{-} \,, \quad 
h^{x}_{N} = {1\over 2} \sinh \eta  \csch \alpha_{+} \,.
\ee
For $\alpha_{\pm} \approx i\pi/2$ or $-i\pi/2$ (i.e., $a_{\pm} \approx
\nu/2$ or $-\nu/2$), the boundary magnetic fields in the $x$ direction are
small and parallel.  Hence, the ground state and corresponding Bethe roots are
``simple''.  Outside of this region of parameter space, the boundary
fields in the $x$ direction are large and/or antiparallel, and so the ground state and
corresponding Bethe roots are ``complicated''.

\section{Discussion}\label{sec:discuss}

We have investigated the ground state of the open XXZ spin chain with
nondiagonal boundary terms which are parametrized by pairs of boundary
parameters, in the thermodynamic limit, using the new exact solutions
\cite{MN1, MN2} and the string hypothesis.  This investigation has
revealed some surprises.  Indeed, for Case I (\ref{caseI}), the ground
state is described in part by a sea of strings of length 2
(\ref{stringhypothesisI}), which is characteristic of spin-1 chains
\cite{spins}.  For Case II (\ref{caseII}), the energy depends on two
sets of Bethe roots (\ref{energyII1}), and in fact on the sum of the
corresponding root densities (\ref{energyII2}).  For each case, there
is a remarkable cancellation of the energy contributions from non-sea
Bethe roots.

Perhaps the biggest surprise is that, for the two cases studied here,
the boundary energies coincide with the result (\ref{ANboundenergy})
for the constrained case (\ref{constraint}), even when that constraint
is not satisfied.  This suggests that the result (\ref{ANboundenergy})
may hold for general values of the boundary parameters. A first step
toward checking this conjecture would be to extend our analysis to the
``unshaded'' regions of parameter space, where the ground state has Bethe 
roots whose imaginary parts depend on the parameters.

We have not computed the Casimir (order $1/N$) energy for the two
cases (\ref{caseI}), (\ref{caseII}).  The computation should be
particularly challenging for the former case, due to the presence of a
complex sea. 
It would be interesting to investigate excited states, and also
applications to other problems, including the boundary sine-Gordon
model. We hope to be able to address such questions in the future.

\section*{Acknowledgments}

This work was supported in part by the National Science Foundation
under Grant PHY-0244261.
 
\begin{appendix}

\section{Appendix}

Here we present more explicit expressions for the bulk and boundary energies.

\subsection{Case I: $p$ even}

The integral appearing in the bulk energy (\ref{bulkenergy}) for $p$ 
even is given 
by
\be
I_{1} &\equiv &
\int_{-\infty}^{\infty}d\lambda\ a_{1}(\lambda)\ s(\lambda) 
={1\over 2\pi} \int_{-\infty}^{\infty}d\omega\ \hat a_{1}(\omega)\
\hat s(\omega) \non \\
&=&  {\mu\over \pi}\sum_{j=1}^{p\over 2}(-1)^{j+{p\over 2}} 
\tan \Big(  \big(j-{1\over 2}\big) \mu\Big) \,.
\label{int1I}
\ee
The parameter-dependent integral appearing in the
boundary energy (\ref{boundenergyeachI}) is given 
by
\be
I_{2}(x) &\equiv & 
{1\over 2\pi} \int_{-\infty}^{\infty}d\omega\ 
{\sinh(\omega/2) \cos (x\omega) \over 
2\sinh(\nu \omega/2)\cosh (\omega/ 2)} \non \\
&=& \sum_{j=1}^{p\over 2}(-1)^{j+{p\over 2}}
b_{j-{1\over 2}}(x/2) - {1\over 2}b_{p+1\over 2}(x/2)\,,
\label{int2I}
\ee
where the function $b_{n}(\lambda)$ is defined in (\ref{anbn}).
Moreover,
\be
{1\over 2\pi} \int_{-\infty}^{\infty}d\omega\ 
{\sinh(\omega/2) \cosh ((\nu-2)\omega/2) \over 
2\sinh(\nu \omega/2)\cosh (\omega/ 2)} 
= I_{2}(i(p-1)/2) \,,
\ee
and
\be
{1\over 2\pi} \int_{-\infty}^{\infty}d\omega\ 
{\sinh((\nu-2)\omega/4) \over 4\sinh(\nu \omega/4)\cosh (\omega/ 2)}
= {1\over 2}\left[ I_{1} -I_{2}(0) \right] \,.
\ee
It follows that the boundary energy (\ref{boundenergyeachI}) is given 
by
\be
E_{boundary}^{\pm} = - {\pi\sin \mu\over \mu} 
\Big\{ {1\over 2}I_{1} - {1\over 2}I_{2}(0) + I_{2}(i(p-1)/2) + 
I_{2}(b_{\pm}) -{1\over 4}\Big\} -{1\over 4}\cos \mu \,,
\ee
where $I_{1}$ and $I_{2}(x)$ are given by (\ref{int1I}) and  (\ref{int2I}), 
respectively.

\subsection{Case II: $p$ odd}

The integral appearing in the bulk energy (\ref{bulkenergy}) for $p$ 
odd is given by
\be
I_{1} &\equiv&
\int_{-\infty}^{\infty}d\lambda\ a_{1}(\lambda)\ s(\lambda) 
={1\over 2\pi} \int_{-\infty}^{\infty}d\omega\ \hat a_{1}(\omega)\
\hat s(\omega) \non \\
&=& {\mu\over \pi^{2}}\Big[ 1  +  2\mu 
\sum_{j=1}^{p-1\over 2}j \cot (j \mu ) \Big] \,.
\label{int1II}
\ee
The parameter-dependent
integral appearing in the boundary energy (\ref{boundenergyeachII}) is
given by
\be
I_{2}(x) &\equiv&
{1\over 2\pi} \int_{-\infty}^{\infty}d\omega\ 
{\sinh(\omega/2) \cosh (x\omega) \over 
2\sinh(\nu \omega/2)\cosh (\omega/ 2)} \non \\
&=&{(-1)^{p-1\over 2}\mu \over \pi \sin(x \pi)}
\left[x + \sum_{j=1}^{p-1\over 2} (-1)^{j} \cot (j \mu ) 
\sin (2 x j \mu) \right] \,.
\label{int2II}
\ee
Moreover,
\be
{1\over 2\pi} \int_{-\infty}^{\infty}d\omega\ 
{\cosh((\nu-2)\omega/4) \over 4\cosh(\nu \omega/4)\cosh (\omega/ 2)}
= {1\over 2}\left[ I_{1} + I_{2}(0) \right] \,.
\ee
It follows that the boundary energy (\ref{boundenergyeachII}) is given 
by
\be
E_{boundary}^{\pm} = - {\pi\sin \mu\over \mu} 
\Big\{ {1\over 2}I_{1} + {1\over 2}I_{2}(0)  + 
I_{2}((p+1-2|a_{\pm}|)/2) -{1\over 4}\Big\} -{1\over 4}\cos \mu \,,
\ee
where $I_{1}$ and $I_{2}(x)$ are given by (\ref{int1II}) and  
(\ref{int2II}), respectively.

\end{appendix}


\begin{thebibliography}{99}
    
\bibitem{Ga}
M. Gaudin,
{\it Phys.  Rev.} {\bf A4}, 386 (1971); {\it La fonction d'onde de
Bethe} (Masson, 1983).

\bibitem{ABBBQ}
F.C. Alcaraz, M.N. Barber, M.T. Batchelor, R.J. Baxter and G.R.W. Quispel,
{\it J. Phys.} {\bf A20},  6397 (1987). 

\bibitem{Sk}
E.K. Sklyanin, 
{\it J. Phys.} {\bf A21}, 2375 (1988).
    
\bibitem{dVGR}
H.J. de Vega and A. Gonz\'alez-Ruiz, 
{\it J. Phys.} {\bf A26}, L519 (1993) 
[{\tt hep-th/9211114}]

\bibitem{GZ}
S. Ghoshal and A. B. Zamolodchikov, 
{\it Int. J. Mod. Phys.} {\bf A9}, 3841 (1994) 
[{\tt hep-th/9306002}]

\bibitem{CLSW}
J. Cao, H.-Q. Lin, K.-J. Shi and Y. Wang, 
[{\tt cond-mat/0212163}]; {\it Nucl. Phys.} {\bf B663}, 487 (2003).

\bibitem{Ne} 
R.I.~Nepomechie, 
{\it J. Stat. Phys.} {\bf 111}, 1363 (2003) 
[{\tt hep-th/0211001}];
R.I.~Nepomechie, 
{\it J. Phys.} {\bf A37},  433 (2004) 
[{\tt hep-th/0304092}]

\bibitem{NR}
R.I.~Nepomechie and F. Ravanini,
{\it J. Phys.} {\bf A36}, 11391 (2003);
Addendum, {\it J. Phys.} {\bf A37}, 1945 (2004)
[{\tt hep-th/0307095}]

\bibitem{YNZ}
W.-L. Yang, R.I.~Nepomechie and Y.-Z. Zhang,
[{\tt hep-th/0511134}]

\bibitem{AN}
C. Ahn and R.I. Nepomechie, 
{\it Nucl. Phys.} {\bf B676},  637 (2004)
[{\tt hep-th/0309261}];

\bibitem{ABNPT}
C. Ahn, Z. Bajnok, R.I. Nepomechie, L. Palla and G. Tak\'acs,  
{\it Nucl. Phys.} {\bf B714}, 307 (2005)
[{\tt hep-th/0501047}]

\bibitem{TBA}
J.-S. Caux, H. Saleur and F. Siano, 
{\it Phys. Rev. Lett.} {\bf 88}, 106402 (2002) 
[{\tt hep-th/0109103}]; 
J.-S. Caux, H. Saleur and F. Siano, 
{\it Nucl. Phys.} {\bf B672}, 411 (2003)
[{\tt hep-th/0306328}];
T. Lee and C. Rim, 
{\it Nucl. Phys.} {\bf B672},  487 (2003)
[{\tt hep-th/0301075}]

\bibitem{generalizations}
A. Doikou, 
{\it Nucl. Phys.} {\bf B668}, 447 (2003)
[{\tt hep-th/0303205}];
J. de Gier and P. Pyatov, 
{\it J. Stat. Mech.} {\bf P03002} (2004)  
[{\tt hep-th/0312235}];
W. Galleas and M.J. Martins, 
{\it Phys. Lett.} {\bf A335}, 167 (2005)
[{\tt nlin.SI/0407027}];
C.S. Melo, G.A.P. Ribeiro and M.J. Martins, 
{\it Nucl. Phys.} {\bf B711}, 565 (2005)
[{\tt nlin.SI/0411038}];
W.-L. Yang, Y.-Z. Zhang and M. Gould, 
{\it Nucl. Phys.} {\bf B698}, 503 (2004) 
[{\tt hep-th/0411048}];
W.-L. Yang and Y.-Z. Zhang, 
{\it JHEP} {\bf 0501}, 021 (2005)
[{\tt hep-th/0411190}];
W.-L. Yang, Y.-Z. Zhang and R. Sasaki, 
{\it Nucl. Phys.} {\bf B729}, 594 (2005)
[{\tt hep-th/0507148}];
J. de Gier, A. Nichols, P. Pyatov and V. Rittenberg,
{\it Nucl. Phys.} {\bf B729}, 387 (2005)
[{\tt hep-th/0505062}];
J. de Gier and F.H.L. Essler,
[{\tt cond-mat/0508707}]

\bibitem{MN1}
R. Murgan and R.I. Nepomechie, 
{\it J. Stat. Mech.} {\bf P05007} (2005);
Addendum, {\it J. Stat. Mech.} {\bf P11004} (2005)
[{\tt hep-th/0504124}]
    
\bibitem{MN2}
R. Murgan and R.I. Nepomechie, 
{\it J. Stat. Mech.} {\bf P08002} (2005)
[{\tt hep-th/0507139}]

\bibitem{Ba}
R.J. Baxter, {\it Exactly Solved Models in Statistical Mechanics}
(Academic Press, 1982).

\bibitem{HQB}
C.J. Hamer, G.R.W. Quispel and M.T. Batchelor, 
{\it J. Phys.} {\bf A20}, 5677 (1987).

\bibitem{Ne01} 
R.I. Nepomechie, 
{\it J. Phys.} {\bf A34}, 9993 (2001)
[{\tt hep-th/0110081}]

\bibitem{YY}
C.N. Yang and C.P. Yang, 
{\it Phys. Rev.} {\bf 150}, 327 (1966).

\bibitem{spins}
H.M. Babujian, 
{\it Nucl. Phys.} {\bf B215}, 317 (1983);
L.A. Takhtajan, 
{\it Phys. Lett.} {\bf 87A}, 479 (1982).


\end{thebibliography}
\end{document}